\documentclass{PoS}
\definecolor{green}{rgb}{0.2, 0.7, 0.3}

\def\blue{\color{blue}}

\def\black{\color{black}}
\def\half{{1\over 2}}

\def\Z{{\mathchoice {\hbox{$\sf\textstyle Z\kern-0.4em Z$}}
{\hbox{$\sf\textstyle Z\kern-0.4em Z$}}
{\hbox{$\sf\scriptstyle Z\kern-0.3em Z$}}
{\hbox{$\sf\scriptscriptstyle Z\kern-0.2em Z$}}}}

\def\square{\kern1pt\vbox{\hrule height 1.2pt\hbox{\vrule width 1.2pt
   \hskip 3pt\vbox{\vskip 6pt}\hskip 3pt\vrule width 0.6pt}
   \hrule height 0.6pt}\kern1pt}
      \def\boxop{{\raise-.25ex\hbox{\square}}}

\def\mn{{\mu\nu}}

\def\tr{{\rm tr}\,}

\def\non{\nonumber\\}

\def\be{\begin{equation}}
\def\ee{\end{equation}\noindent}
\def\bear{\begin{eqnarray}}
\def\ear{\end{eqnarray}\noindent}
\def\bec{\blue\begin{equation}}
\def\eec{\end{equation}\black\noindent}
\def\bearc{\blue\begin{eqnarray}}
\def\earc{\end{eqnarray}\black\noindent}
\def\benn{\begin{enumerate}}
\def\enn{\end{enumerate}}

\def\slash#1{#1\!\!\!\raise.15ex\hbox {/}}
\newcommand{\slD}{\,\raise.15ex\hbox{$/$}\kern-.27em\hbox{$\!\!\!D$}}
\newcommand{\slpartial}{\raise.15ex\hbox{$/$}\kern-.57em\hbox{$\partial$}}

\newcommand{\nc}{\newcommand}
\nc{\spa}[3]{\left\langle#1\,#3\right\rangle}
\nc{\spb}[3]{\left[#1\,#3\right]}
\nc{\ksl}{\not{\hbox{\kern-2.3pt $k$}}}
\nc{\hf}{\textstyle{1\over2}}
\nc{\pol}{\varepsilon}
\nc{\tq}{{\tilde q}}
\nc{\esl}{\not{\hbox{\kern-2.3pt $\pol$}}}

\def\kinb{{1\over 4}\dot x^2}

\def\epseps#1#2{\varepsilon_{#1}\cdot \varepsilon_{#2}}
\def\epsk#1#2{\varepsilon_{#1}\cdot p_{#2}}
\def\kk#1#2{p_{#1}\cdot p_{#2}}

\def\4piTD{{(4\pi T)}^{-{D\over 2}}}
\def\4piT4{{(4\pi T)}^{-2}}

\def\Tintm4{{\dps\int_{0}^{\infty}}{dT\over T}\,e^{-m^2T}
    {(4\pi T)}^{-2}}
\def\Tintm{{\dps\int_{0}^{\infty}}{dT\over T}\,e^{-m^2T}}

\newcommand{\slG}{{{\dot G}\!\!\!\! \raise.15ex\hbox {/}}}
\newcommand{\Gd}{{\dot G}}

\def\GBd12{{\dot G}_{B12}}

\newcommand{\no}{\noindent}
\def\non{\nonumber}
\def\dps{\displaystyle}

\title{Form factor decomposition of the off-shell four-gluon amplitudes }

\ShortTitle{Form factor decomposition of the off-shell four-gluon amplitudes }

\author {\speaker{Naser Ahmadiniaz} and {Christian Schubert}\\

{ Instituto de F\'{\i}sica y Matem\'aticas
\\
Universidad Michoacana de San Nicol\'as de Hidalgo\\
Edificio C-3, Apdo. Postal 2-82\\
C.P. 58040, Morelia, Michoac\'an, M\'exico}\\

        E-mail: \email{naser@ifm.umich.mx, schubert@ifm.umich.mx}}

\abstract{We show how to use the Bern-Kosower master formula, originally a generating functional for 
on-shell gluon matrix elements, to derive well-organized form factor decompositions of the
off-shell one-particle-irreducible N - gluon vertices. Two such algorithms are presented which can be used for any N, 
the first one optimized with respect to the nonabelian gauge invariance, the second one with respect to transversality. We give
explicit results for the three- and four-gluon cases. The second algorithm in the three-point case reproduces precisely the
well-known Ball-Chiu decomposition, and in the four-point case a natural generalization thereof. A particularly
simple structure emerges in the ${\cal N}=4$ SYM case.}

\FullConference{QCD-TNT-III-From quarks and gluons to hadronic matter: A bridge too far?,\\
		2-6 September, 2013\\
		European Centre for Theoretical Studies in Nuclear Physics and Related Areas (ECT*), Villazzano, Trento (Italy)}

\begin{document}

\section{Introduction}

Multi-gluon amplitudes in perturbative QCD pose two very different computational challenges. On one hand, there
is the calculation of on-shell matrix elements, a field that has seen tremendous activity and progress
during the last decade, particularly so for the massless and/or SUSY cases. 
A whole host of new concepts and techniques have emerged, such as 
unitarity-based methods \cite{bddk-unitarity,berhua}, twistors \cite{witten-twistors}, BCFW recursion \cite{brcafe,bcfw},
and Grassmannians \cite{acck,masski}; see \cite{elvhua-review} and \cite{dixon-reviewnew} for recent reviews.

On the other hand, there are the off-shell gluon amplitudes, which carry additional physical information, particularly
on the infrared properties of QCD (see, e.g., \cite{alhusc}). 
They are also essential for the matching of perturbative information with lattice data (see, e.g., \cite{petiws}),
and constitute an essential ingredient for the Schwinger-Dyson equations. In the latter context, previously only the two and three-point
amplitudes were used with their full loop-corrected structure, but a need for the inclusion of the four-gluon
vertex is already felt \cite{madalk}.
Off-shell, the natural building blocks are not the connected but the one-particle-irreducible  (`1PI') $N$ - gluon
Green's functions, also called `$N$ - gluon vertices'.  For their calculation, no progress comparable to the on-shell case 
was achieved in recent years, and their explicit computation is presently essentially still stuck at the three-point level. 

Let us shortly review what is known about the $N$ - gluon vertices. Following early work by Celmaster and Gonsalves \cite{celgon}
and Pascual and Tarrach \cite{pastar}, in 1980 Ball and Chiu  \cite{balchi2}
studied the off-shell gluon amplitudes for the gluon loop  in Feynman gauge.
They analyzed the Ward identities and derived a form factor (``Ball-Chiu'') decomposition of the three-gluon vertex, valid to all loop orders.
Cornwall and Papavassiliou \cite{corpap} in 1989 constructed a ``gauge invariant three-gluon vertex''  through the pinch technique 
(see \cite{binpap-rev} for a review of this technique).
Freedman et al. in 1992 studied the conformal properties of this vertex \cite{fgjr}.
In 1993 J. Papavassiliou \cite{papavassiliou-4gluon} extended this construction to the four-gluon vertex.
Davydychev, Osland and Tarasov \cite{daosta-3gluonD} in 1996 
calculated the gluon loop contribution to the one-loop three-gluon vertex
in arbitrary covariant gauge, and also the massless fermion loop contribution.
The fermion loop calculation was later generalized to the massive case by 
Davydychev, Osland and Saks \cite{daossa-3gluonDm}. 
In 2006, Binger and Brodsky \cite{binbro}
studied the one-loop three-gluon vertex in various dimensions, 
using the background field method \cite{abbott,abgrsc}. 
Besides the gluon and fermion loop cases, they now also included the scalar loop,
as is needed for SUSY extensions of QCD, and they derived various sum rules relevant
to the SUSY case.
They also verified that, as had been suggested in \cite{dewedi,papavassiliou95}, 
in the gluon loop case the use of the background field method with
quantum Feynman gauge leads to precisely the same vertex as the pinch technique.
At two loops, the three-gluon vertex has been
obtained so far only for some very special momentum configurations \cite{daosta-2loop,davosl,gracey}). 

Our aim here is to demonstrate that the string-inspired approach to perturbative QCD,
originally developed in the on-shell context by Bern and Kosower \cite{berkos:prl,berkos:npb362,berkos:npb379}, 
is also extremely promising as a tool for the derivation of form factor decompositions of 
the $N$ - gluon amplitudes off-shell (see also J. Cornwall's talk at this workshop \cite{cornwall-trento}).

\section{The Ward identities for the off-shell gluon amplitudes}

Off-shell, the Ward identities for the gluon amplitudes are inhomogeneous and map
the $N$ - vertex to $N-1$ - point amplitudes. 
E.g. for the four-point case one finds \cite{papavassiliou-4gluon}

\bear
p_1^{\mu}\Gamma_{\mu\nu\alpha\beta}^{abcd} 
(p_1,p_2,p_3,p_4)
= f_{abe}\Gamma_{\nu\alpha\beta}^{cde}(p_1+p_2,p_3,p_4) + \, {\rm perm.}
\ear
These identities as they stand hold for the scalar and spinor loop unambiguously, but 
for the gluon loop only if one uses the pinch technique, or equivalently the background field method with quantum Feynman gauge. 
Other gauge fixings of the gluon loop will generally lead to a more complicated right-hand side involving ghosts.

\section{The QCD three-gluon vertex and its Ball-Chiu decomposition}

\begin{figure}[h]
\hspace{100pt}{\centering
\includegraphics{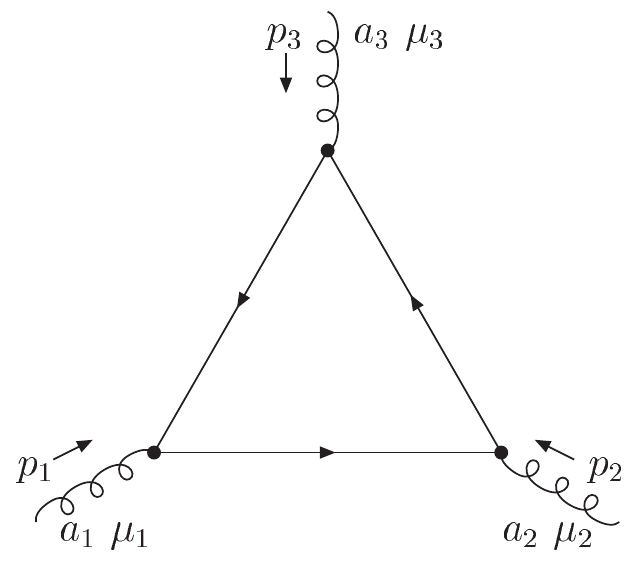}
}
\caption{Three-gluon vertex.}
\label{fig1}
\end{figure}

\no
The three-gluon vertex in QCD at tree level,

\bear
-igf^{a_1a_2a_3}\bigl[g_{\mu_1\mu_2}(p_1-p_2)_{\mu_3} + cycl. \bigr]\,,
\label{3gltree}
\ear
is corrected at the  one-loop level by the 1PI  three-gluon vertex  with a spinor or gluon loop.
E.g. for the spinor loop case we have the diagram shown in Fig. \ref{fig1} (and a second one with the
other orientation of the fermion).

\noindent
The Ball-Chiu decomposition of the vertex is \cite{balchi2}

\begin{eqnarray}
\Gamma_{\mu_{1}\mu_{2}\mu_{3}}(p_{1},p_{2},p_{3})&=&
A(p_{1}^2,p_{2}^2;p_{3}^2) g_{\mu_{1}\mu_{2}}(p_{1}-p_{2})_{\mu_{3}}
+ B(p_{1}^2,p_{2}^2;p_{3}^2)
g_{\mu_{1}\mu_{2}}(p_{1}+p_{2})_{\mu_{3}}\nonumber\\ 
&&\hspace{-20pt} 
+ C(p_{1}^2,p_{2}^2;p_{3}^2)
 [p_{1\mu_{2}}p_{2\mu_{1}}-p_{1}\cdot p_{2}g_{\mu_{1}\mu_{2}}](p_{1}-p_{2})_{\mu_{3}}\nonumber\\&&\hspace{-20pt} 
+\frac{1}{3} S(p_{1}^2,p_{2}^2,p_{3}^2)(p_{1\mu_{3}}p_{2\mu_{1}}p_{3\mu_{2}}+p_{1\mu_{2}}p_{2\mu_{3}}p_{3\mu_{1}})
\nonumber\\ &&\hspace{-20pt}
+ F(p_{1}^2,p_{2}^2;p_{3}^2)
[p_{1\mu_{2}}p_{2\mu_{1}}-p_{1}\cdot p_{2}g_{\mu_{1}\mu_{2}}]
[p_{2\mu_{3}}p_{1}\cdot p_{3}-p_{1\mu_{3}}p_{2}\cdot p_{3}]\nonumber\\&&\hspace{-20pt} +
 H(p_{1}^2,p_{2}^2,p_{3}^2)
 \Bigl(-g_{\mu_{1}\mu_{2}}[p_{1\mu_{3}}p_{2}\cdot p_{3}-p_{2\mu_{3}}p_{1}\cdot p_{3}]
+\frac{1}{3}(p_{1\mu_{3}}p_{2\mu_{1}}p_{3\mu_{2}}-p_{1\mu_{2}}p_{2\mu_{3}}p_{3\mu_{1}})\Bigr)\nonumber\\
&&\hspace{-20pt} + \, [ \mbox{\rm cyclic permutations of }(p_{1},\mu_{1}), (p_{2},\mu_{2}),(p_{3},\mu_{3})]\,.\nonumber\\
\end{eqnarray}
This form factor decomposition is universal, that is, 
valid for the scalar, spinor and gluon loop, and also for higher loop corrections. 
Only the coefficient functions $A,B,C,F,H,S$ change. 
At tree level, $A=1$, the other functions vanish. Explicit calculation shows that
$S$ still vanishes at one-loop. 
The tensor structures multiplying $F,H$ are manifestly transversal.

\section{The string-inspired formalism}

In 1991 Bern and Kosower in their seminal work derived, by an analysis of the infinite string limit of 
certain string amplitudes, the following Bern-Kosower master formula 
\cite{berkos:prl,berkos:npb362,berkos:npb379}:

\bear
 \Gamma^{a_{1}\dots a_{N}}[p_{1},\varepsilon_{1};\dots;p_{N},\varepsilon_{N}]
 &=&(-ig)^{N}\mbox{tr}(T^{a_{1}}\dots T^{a_{N}})
 \int_{0}^{\infty} dT(4\pi T)^{-D/2}e^{-m^2 T}\nonumber\\
 && \times\int_{0}^{T}d\tau_{1}\int_0^{\tau_{1}}d\tau_2\dots\int_0^{\tau_{N-2}}d\tau_{N-1}
 \nonumber\\ && \times
 \exp\Bigg\{\sum_{i,j=1}^N\left[\frac{1}{2}
  G_{Bij}p_{i}\cdot p_{j}
-i\dot{G}_{Bij}\varepsilon_{i}\cdot p_{j}+\frac{1}{2}\ddot{G}_{Bij}\varepsilon_{i}\cdot\varepsilon_{j}\right]\Bigg\}
\Biggl\vert_{\rm lin (\varepsilon_1 \ldots \varepsilon_N)} \,.
\nonumber\\
\ear
As it stands, this is a parameter integral representation for the (color-ordered) 1PI $N$ - gluon amplitude  induced by a {scalar} loop, with momenta
$p_i$ and polarizations $\varepsilon_i$, in $D$ dimensions.
Here $m$ and $T$ are the loop mass and proper-time, $\tau_i$ fixes the location of the $i$th gluon,
and $G_{Bij}\equiv G_B(\tau_i,\tau_j)$ denotes the ``bosonic''  worldline Green's function, defined by

\bear
G_{Bij}&=& |\tau_i - \tau_j | - \frac{(\tau_i - \tau_j)^2}{T}\,,
\ear
and dots generally denote a derivative acting on the first variable. Explicitly,
\bear
 \dot G_B(\tau_1,\tau_2)&=& {\rm sign}(\tau_1 - \tau_2)
- 2 {{(\tau_1 - \tau_2)}\over T}\, \nonumber\\
\ddot G_B(\tau_1,\tau_2)
&=& 2 {\delta}(\tau_1 - \tau_2)
- {2\over T} \,.
\nonumber\\
\label{GB}
 \ear
In the Bern-Kosower formalism, the master formula serves as a generating
functional for the full on-shell $N$ - gluon amplitudes for the scalar, spinor and gluon loop,
through the use of the {\it Bern-Kosower rules}:

\begin{itemize}
\item
For fixed $N$, expand the generating exponential and take only the terms linear in all polarization vectors.

\item
Use suitable integrations-by-parts (IBPs) 
to remove all second derivatives  $\ddot G_{Bij}$.

\item
Apply two types of pattern-matching rules:
\begin{itemize}
\item
The ``tree replacement rules'' 
generate (from a field theory point of view) the contributions of the missing reducible diagrams.

\item
The ``loop replacement rules''  generate 
the integrands for the spinor and gluon loop from the one 
for the scalar loop.
\end{itemize}

\end{itemize}

\section{The worldline path integral approach}

Shortly after the work of Bern and Kosower, Strassler \cite{strassler} 
rederived the master formula and the loop replacement rules
using worldline path integral representations of the gluonic effective actions.
E.g. for the scalar loop, 

\be 
\Gamma [A]
=
{\rm tr}
\Tintm
\int{\cal D}x(\tau)\,
{\cal P}
e^{-\int_0^T d\tau\Bigl(
\kinb
+ig\dot x\cdot A(x(\tau))
\Bigr)}\,,
\nonumber\\
\ee\no
where $A_{\mu}=A_{\mu}^aT^a$ and $\cal P$ denotes path ordering.
This also showed that the master formula and the loop replacement rules hold off-shell, 
which was not obvious from its original derivation.
Moreover, in a beautiful but unpublished paper \cite{strassler2}
Strassler noted that the IBP generates automatically abelian field strength tensors

\bear
f_i^{\mu\nu} \equiv
p_i^{\mu}\varepsilon_i^{\nu}
- \varepsilon_i^{\mu}p_i^{\nu}
\label{deff}\,,
\ear
in the bulk, and
color commutators $ [T^{a_i},T^{a_j}]$ as boundary terms.
Those fit together to produce full nonabelian field strength tensors

\bear
F_{\mn} \equiv F_{\mn}^a T^a = (\partial_{\mu}A_{\nu}^a - \partial_{\nu}A_{\mu}^a) T^a + ig[A_{\mu}^bT^b,A_{\nu}^cT^c]
\label{defF}\,,
\ear
in the low-energy effective action.
Thus we see something very interesting, namely 
the emergence of gauge invariant tensor structures {\it at the integrand level}.

However, the removal of all $\ddot G_{Bij}$ by IBP can be done in many ways! 
Moreover, it is not obvious how to do it without breaking the bose symmetry between
the gluons. 
In \cite{strassler2} Strassler started to investigate this ambiguity at the four-point level,
but an algorithm valid for any $N$ and manifestly preserving the permutation
invariance was given only much later \cite{26,41} by one of the authors.
This algorithm still followed the objective of achieving a form of the $N$ - gluon vertex that,
in $x$ - space, would correspond to a manifestly covariant representation of the
nonabelian effective action. However, it turns out not be optimized from another point
of view, which is important, e.g., for the Schwinger-Dyson equations, namely it does
not lead to a clean separation of the vertices into transversal and longitudinal parts.
This remaining obstacle has been overcome only recently in \cite{91}, where we
give two IBP algorithms that work for arbitrary $N$ and
 lead to explicit form-factor decompositions of the off-shell $N$ - gluon amplitudes: 

\begin{itemize}

\item
The first algorithm uses only local total derivative terms and leads to a 
representation that matches term-by-term with the low-energy effective action
(``Q-representation '').

\item
The second algorithm uses both local and nonlocal total derivative terms and 
leads to the transversality of all bulk terms at the integrand level (`` S-representation'').

\end{itemize}

\no
In \cite{92}, we applied both algorithms to the three-point case and showed that, in particular,  
the second algorithm generates precisely the Ball-Chiu decomposition.
Very recently, we have carried out the same program also for the four-gluon vertex \cite{4gluon}.
We will now sketch these rather involved calculations as well as space permits.

\section{The Q-representation of the three-gluon vertex}

For $N = 3$, the Q-representation is (for the scalar loop) \cite{92}

\bear
\Gamma &=& 
\frac{g^3}{(4\pi)^{\frac{D}{2}}}\mbox{tr}(T^{a_{1}} [T^{a_{2}},T^{a_{3}}])(\Gamma^{3} 
+ \Gamma^{2} + \Gamma^{{\rm bt}})\,,
\nonumber\\
\ear
where

\bear
\Gamma^{3} &=& - \int_{0}^{\infty} \frac{dT}{T^{\frac{D}{2}}}e^{-m^2 T}\int_{0}^{T}d\tau_{1}\int_0^{\tau_{1}}d\tau_{2}\, Q_3^3 
\exp\bigg\{\sum_{i,j=1}^3\frac{1}{2}G_{Bij}p_{i}\cdot p_{j} \bigg\}
\nonumber\\
\Gamma^{2} &=& \Gamma^{3}(Q_3^3\to Q_3^2) \nonumber\\
\Gamma^{{\rm bt}} &=&
\int_{0}^{\infty} \frac{dT}{T^{\frac{D}{2}}}e^{-m^2 T}\int_{0}^{T}d\tau_{1}
\dot{G}_{B12}\dot{G}_{B21} 
\Bigl\lbrack\varepsilon_3\cdot f_1\cdot\varepsilon_2
\,e^{G_{B12}p_{1}\cdot (p_{2}+p_{3})} +
{\rm cycl.}
\Bigr\rbrack\,, \nonumber\\
\ear
and

\bear
Q_{3}^3&=&\dot{G}_{B12}\dot{G}_{B23}\dot{G}_{B31}\tr(f_1f_2f_3) \nonumber\\
Q_{3}^2&=& \half \dot{G}_{B12}\dot{G}_{B21}\tr (f_1f_2)\dot{G}_{B3k}\varepsilon_{3}\cdot p_{k}+
2 \, {\rm perm}\,.
\nonumber\\
\label{Q3}
\ear
Here $\Gamma^{\rm bt}$ comes from the boundary terms, and the upper indices on 
$\Gamma^{2,3},Q^{2,3}$ refer to the ``cycle content''; e.g. $Q_3^3$ contains a factor
$\dot{G}_{B12}\dot{G}_{B23}\dot{G}_{B31}$ whose indices form a closed cycle involving
three points, called ``three-cycle''.
Dummy indices like the one appearing in $Q_3^2$ are to be summed from $1$ to $N=3$. 
To pass from the scalar to the spinor loop, one applies the ``loop replacement rules''

\bear
\dot G_{Bij}\dot G_{Bji} &\to & \dot G_{Bij}\dot G_{Bji} - G_{Fij}G_{Fji} \nonumber\\
\dot G_{B12}\dot G_{B23}\dot G_{B31} &\to & \dot G_{B12}\dot G_{B23}\dot G_{B31} - G_{F12}G_{F23}G_{F31}\,, \nonumber\\
\label{spin}
\ear
where $G_{Fij} = {\rm sign}(\tau_i-\tau_j)$.
Similarly, the integrand for the gluon loop is obtained from the scalar loop one by

\bear
\dot G_{Bij}\dot G_{Bji} &\to & \dot G_{Bij}\dot G_{Bji} - 4G_{Fij}G_{Fji} \nonumber\\
\dot G_{B12}\dot G_{B23}\dot G_{B31} &\to & \dot G_{B12}\dot G_{B23}\dot G_{B31} - 4G_{F12}G_{F23}G_{F31}\,. \nonumber\\
\label{gluon}
\ear
As stated above, the gluon loop vertex obtained in this way corresponds to the background field method with quantum Feynman gauge
\cite{strassler,18}. And for all three cases - scalar, spinor and gluon loop - the vertex allows a perfect match with the low-energy effective action.
We recall that the low energy expansion of the one-loop QCD effective action induced by a
loop particle of mass $m$ has the form (see, e.g., \cite{25})

\begin{equation}
\Gamma [F] = \int_0^\infty \!{dT\over T} \; 
\frac{{\rm e}^{-m^2 T}}{(4\pi T)^{D/2}} \; 
{\rm tr} \; \int \! dx_0 \; \sum_{n=2}^{\infty} \; 
\frac{(-T)^n}{n!} \; O_n[F] \,,
\nonumber
\end{equation}\no
where $O_n(F)$is a Lorentz and gauge invariant expression of mass dimension $2n$.
To lowest orders,

\begin{eqnarray}
O_2 &=& c_2 g^2 F_{\mu\nu} F_{\mu\nu} \, ,\nonumber\\
O_3 &=& 
          c_3^3\,i g^3\,F_{\kappa\lambda}F_{\lambda\mu}F_{\mu\kappa} 
          + c_3^2 g^2D_{\lambda}F_{\mu\nu}D^{\lambda}F^{\mu\nu} \,,
          \nonumber\\
\end{eqnarray}
where only the coefficients $c_2,c_3^{2,3}$ depend on the spin of the loop particle. 
We recognize the correspondences

\begin{eqnarray}
&&\Gamma^3\leftrightarrow F_{\kappa}^{~\lambda}F_{\lambda}^{~\mu}F_{\mu}^{~\kappa}
=f_{\kappa}^\lambda f_{\lambda}^\mu f_{\mu}^\kappa+{\rm higher~point~terms}\nonumber\\
&&\Gamma^2\leftrightarrow (\partial+ig\underbrace {A)F(\partial}+igA)F\nonumber\\
&&\Gamma^{{\rm bt}}\leftrightarrow(f+ig\underbrace{[A,A])(f}+ig[A,A])\,.\nonumber\\
\end{eqnarray}

\section{The S-representation of the three-gluon vertex}

\no
In the S-representation, the three-gluon vertex becomes

\bear
\tilde\Gamma &=& 
\frac{g^3}{(4\pi)^{\frac{D}{2}}}\mbox{tr}(T^{a_{1}} [T^{a_{2}},T^{a_{3}}])(\tilde\Gamma^{3} 
+ \tilde\Gamma^{2} + \tilde\Gamma^{{\rm bt}})\,,
\nonumber\\
\ear
where

\bear
\tilde\Gamma^{3} &=& - \int_{0}^{\infty} \frac{dT}{T^{\frac{D}{2}}}e^{-m^2 T}\int_{0}^{T}d\tau_{1}\int_0^{\tau_{1}}d\tau_{2}\, S_3^3 
\exp\bigg\{\sum_{i,j=1}^3\frac{1}{2}G_{Bij}p_{i}\cdot p_{j} \biggr \}
\nonumber\\
\tilde\Gamma^{2} &=& \tilde\Gamma^{3}(S_3^3\to S_3^2) \nonumber\\
\tilde\Gamma^{{\rm bt}} &=&
\int_{0}^{\infty} \frac{dT}{T^{\frac{D}{2}}}e^{-m^2 T}\int_{0}^{T}d\tau_{1}
\dot{G}_{B12}\dot{G}_{B21} 
\Bigl\lbrace\Bigl[ \varepsilon_3\cdot f_1\cdot\varepsilon_2
-\half {\rm tr}(f_1f_2)\rho_3+\half {\rm tr}(f_3f_1)\rho_2\Bigr]
\nonumber\\
&& \hspace{140pt}\times \,e^{G_{B12}p_{1}\cdot (p_{2}+p_{3})}  +\, {\rm cycl.}\Bigr\rbrace\,, \nonumber\\
\ear
and

\vspace{-23pt}
\bear
S_{3}^3&=&\dot{G}_{B12}\dot{G}_{B23}\dot{G}_{B31}\tr(f_1f_2f_3) \nonumber\\
S_{3}^2&=& \half \dot{G}_{B12}\dot{G}_{B21}\tr (f_1f_2)\dot{G}_{B3k}\frac{r_3\cdot f_3\cdot p_k}{r_3\cdot p_3}
+ 2 \, {\rm perm}\,.
\nonumber\\
\label{Q3}
\ear
Here we have introduced three vectors $r_i$ which obey $ r_i \cdot p_i \ne 0$ but are arbitrary otherwise,
and $ \rho_i:=\frac{r_i\cdot\varepsilon_i}{r_i\cdot p_i}$.
Note that $S^3$ is the same as $Q^3$ above, but that in $S^2$, contrary to $Q^2$, all three polarization
vectors $\varepsilon_i$ are absorbed in abelian field strength tensors $f_i$.
Thus all bulk terms are now manifestly transversal, even at the integrand level,
and it turns out that with the cyclic choice  $$r_{1}=p_2-p_3, r_{2}=p_3-p_1, r_{3}=p_1-p_2\,,$$ we 
get a term-by-term match with the Ball-Chiu decomposition:

\bear
H(p_1^2,p_2^2,p_3^2)&=&
C(r)\frac{d_0g^2}{(4\pi)^{D/2}}\Gamma(3-\frac{D}{2}) I^{D}_{3,B} (p_1^{2},p_2^{2},p_3^{2}) \nonumber\\
A (p_1^2,p_2^2;p_3^2)&=&
C(r)\frac{d_0g^2}{2(4\pi)^{D/2}}\Gamma(2-\frac{D}{2})\Bigl[I^{D}_{{\rm bt, B}}(p_1^2)+I^{D}_{{\rm bt, B}}(p_2^2)\Bigr] \nonumber\\
B(p_1^2,p_2^2;p_3^2)&=&
C(r)\frac{d_0g^2}{2(4\pi)^{D/2}}
\Gamma(2-\frac{D}{2})\Big[I^{D}_{{\rm bt, B}}(p_1^2)-I^{D}_{{\rm bt, B}}(p_2^2)\Big] \nonumber\\
F(p_1^2,p_2^2;p_3^2)&=&
C(r)\frac{d_0g^2}{(4\pi)^{D/2}}\Gamma(3-\frac{D}{2})
\frac{I_{2,B}^{D}(p_1^2,p_2^2,p_3^2)-I_{2,B}^{D}(p_2^2,p_1^2,p_3^2)}{p_1^2-p_2^2} \nonumber\\
C(p_1^2,p_2^2;p_3^2)&=&
C(r)\frac{d_0g^2}{(4\pi)^{D/2}}\Gamma(2-\frac{D}{2}) \frac{I^{D}_{{\rm bt, B}}(p_1^2)-I^{D}_{{\rm bt, B}}(p_2^2)}{p_1^2-p_2^2}\nonumber\\
S(p_1^2,p_2^2;p_3^2)&=&0\,,\nonumber\\
\ear
where we have used  ${\rm tr}(T^{a_1}[T^{a_2},T^{a_3}])=iC( r)\,f^{a_1a_2a_3}$.\\
Here we have written down the scalar loop case, but due to the loop replacement rules (\ref{spin}) and (\ref{gluon}), the spinor and gluon
loop cases differ from it only in the coefficient functions on the right-hand sides. 
For completeness, let us write down these coefficient functions also in terms of standard Feynman-Schwinger parameter integrals
(again for the scalar loop)

\bear
I_{3,B}^D(p_1^2,p_2^2,p_3^2) &=& \int_0^1d{\alpha}_1d{\alpha}_2d{\alpha}_3\delta(1-{\alpha}_1-{\alpha}_2-{\alpha}_3)
\times\frac{(1-2{\alpha}_1)(1-2{\alpha}_2)(1-2{\alpha}_3)}{\Bigl( m^2 + {\alpha}_1{\alpha}_2p_1^2+{\alpha}_2{\alpha}_3p_2^2+{\alpha}_1{\alpha}_3p_3^2\Bigr)^{3-\frac{D}{2}}}\nonumber\\
I_{2,B}^D(p_1^2,p_2^2,p_3^2) &=& \int_0^1d{\alpha}_1d{\alpha}_2d{\alpha}_3\delta(1-{\alpha}_1-{\alpha}_2-{\alpha}_3)
\times\frac{(1-2{\alpha}_2)^2(1-2{\alpha}_1)}{\Bigl( m^2 + {\alpha}_1{\alpha}_2p_1^2+{\alpha}_2{\alpha}_3p_2^2+{\alpha}_1{\alpha}_3p_3^2\Bigr)^{3-\frac{D}{2}}}\nonumber\\
I_{bt,B}^D(p^2) &=& \int_0^1d{\alpha}\frac{(1-2{\alpha})^2}{\bigl( m^2 + {\alpha}(1-{\alpha})p^2\bigr)^{2-\frac{D}{2}}}\,.\nonumber\\
\ear

\noindent
They are, of course, of the same type as the ones arising in other approaches.

\section{The four-gluon vertex}

\no
Proceeding to the four-point case, here the Q-representation for the scalar loop has the following bulk terms:

\bear
\Gamma^{a_1a_2a_3a_4} &=&  
g^4\mbox{tr}(T^{a_{1}}\dots T^{a_{4}})
 \int_{0}^{\infty} dT(4\pi T)^{-D/2}e^{-m^2 T}\nonumber\\
 && \times\int_{0}^{T}d\tau_{1}\int_0^{\tau_{1}}d\tau_2\int_0^{\tau_2}d\tau_3
 Q_4
\exp\bigg\{\sum_{i,j=1}^4\frac{1}{2} G_{Bij}p_{i}\cdot p_{j}
\bigg\}\,,
\nonumber\\
\ear
\bear
Q_4&=&Q^4_4+Q^3_4+Q^2_4-Q^{22}_{4}\non\\
Q^4_4&=&\Gd(1234)+\Gd(1243)+\Gd(1324)\non\\
Q^3_4&=&\Gd(123)T(4)+\Gd(234)T(1)+\Gd(341)T(2)+\Gd(412)T(3)\non\\
Q^2_4&=&\Gd(12)T(34)+\Gd(13)T(24)+\Gd(14)T(23)+\Gd(23)T(14)\non\\
&&+\Gd(24)T(13)+\Gd(34)T(12)\non\\
Q^{22}_4&=&\Gd(12)\Gd(34)+\Gd(13)\Gd(24)+\Gd(14)\Gd(23)\,,\nonumber\\
\ear
where we have now employed a more condensed notation:

\bear
\Gd(i_1i_2\cdots i_n)&:=&\Gd_{Bi_1i_2}\Gd_{Bi_2i_3}\cdots\Gd_{Bi_ni_1}\Bigl(\half\Bigr)^{\delta_{n,2}} {\rm tr}(f_{i_1}f_{i_2}\cdots f_{i_n})\nonumber\\
T(i)&:=&\sum_r\Gd_{Bir}\epsk ir\nonumber\\
T(ij)&:=&\sum_{r,s}\Bigg\{\Gd_{Bir}\epsk ir\Gd_{js}\epsk js+\half\Gd_{Bij}\epseps ij\Big\lbrack\Gd_{Bir}\kk ir-\Gd_{Bjr}\kk jr\Big\rbrack\Bigg\}\,.\nonumber\\
\ear
The IBP procedure now leads to both single boundary terms (three-point integrals) and double boundary terms (two-point integrals).
The following rules emerge:

\begin{itemize}

\item
Each single boundary term, say for the limit $ 3\to 4$, matches  some bulk term in the Q-representation of the three-gluon vertex,
with momenta $(p_1,p_2,p_3+p_4)$, and $f_3=p_3 \otimes \varepsilon_3- \varepsilon_3\otimes p_3$ 
replaced by $\varepsilon_3\otimes \varepsilon_4 -  \varepsilon_4\otimes \varepsilon_3 $.

\item
Each double boundary term, say for the limit $1\to2 , 3\to 4$, matches  the bulk term in the Q-representation of the 
two-point function,
with momenta $(p_1+p_2,p_3+p_4)$, and the double replacement

\bear
f_1& =&p_1 \otimes \varepsilon_1- \varepsilon_1\otimes p_1
 \to 
 \varepsilon_1\otimes \varepsilon_2 -  \varepsilon_2\otimes \varepsilon_1 
\nonumber\\
 f_2 &= &p_2 \otimes \varepsilon_2- \varepsilon_2\otimes p_2
 \to 
 \varepsilon_3\otimes \varepsilon_4 -  \varepsilon_4\otimes \varepsilon_3 \,.
\nonumber\\
\ear

\end{itemize}

\no
Effectively, a boundary term always completes a $f_i$ to a full nonabelian field strength tensor;
that is, we are seeing just the projection to plane waves of the completion

\bear
\partial_{\mu}A_{\nu} -  \partial_{\nu}A_{\mu}  \to \partial_{\mu}A_{\nu} -  \partial_{\nu}A_{\mu} + ig[A_{\mu},A_{\nu}]\,.
\label{completion}
\ear
Moreover, this recursive structure is compatible with the replacement rules.

The S-representation looks similar, but has the bulk terms written completely in terms of the $f_i$, so that all non-transversality has
now been pushed into the boundary terms. It involves now the
choice of four vectors $r_i$ with $r_i \cdot p_i \ne 0$.

\section{Off-shell one-loop four-gluon vertex in $ {\cal N}=4$ SYM}

In $ {\cal N}=4$ SYM the one-loop two - and three - gluon amplitudes vanish 
(this relates to the conformal invariance and finiteness of the theory). 
The one-loop four-gluon vertex is the first non-vanishing one, and it is
extremely simple: all boundary terms cancel out (since they would covariantize
the nonexisting lower point amplitudes) and the bulk term factors as

\bear
\Gamma^{a_1a_2a_3a_4} &=&
4g^4 \tr (T^{a_1}T^{a_2}T^{a_3}T^{a_4})
F^4_{\rm ss}
 B(1234) \, + \, {\rm non-cyclic \,\, permutations}\,,
 \nonumber\\
\label{Gamma}
\ear
where
$B(1234)$ is the off-shell scalar box integral with momenta $p_1,\ldots, p_4$, and
the  whole Lorentz structure is contained in the invariant

\bear
F^4_{\rm ss}&=& \tr(f_1f_2f_3f_4) +\tr (f_1f_2f_4f_3) + \tr (f_1f_3f_2f_4) \nonumber\\
&& - \frac{1}{4} \tr (f_1f_2)\tr (f_3f_4) - \frac{1}{4} \tr (f_1f_3)\tr (f_2f_4) 
- \frac{1}{4} \tr (f_1f_4) \tr (f_2f_3) \,.
\nonumber\\
\ear
Although this explicit form of the off-shell one-loop four-gluon amplitude in ${\cal N}=4$ SYM appears to be new,
the Lorentz tensor $F_{\rm ss}$ is well-known to string theorists, since it appears in the low energy expansion of the
effective action of the open superstring; see \cite{brodix} and refs therein.

\section{Summary and Outlook}

\no
To summarize, the main points which we wanted to make here are:

\begin{itemize}

\item
In the string-inspired formalism, form factor decompositions of the $N$ - vertex compatible with Bose symmetry
and gauge invariance can be generated simply by an integration-by-parts procedure
starting from the Bern-Kosower master formalism, which originally was derived as a generating functional
for on-shell matrix elements. 

\item
At the one-loop level, the parameter integrals appearing in the form factors for the scalar, spinor and gluon
loop cases are all obtained directly from the Bern-Kosower master formula.

\item
We have carried out this program explicitly for the three- and four-point cases. 

\item
In particular, we have obtained a natural four-point generalization of the Ball-Chiu
decomposition. It is distinguished by the fact that all true four-point terms are manifestly transversal,
so that all longitudinal components are given by lower-point integrals.

\end{itemize}

\end{document}